\documentclass[12pt]{article}
\usepackage[margin=1.9cm]{geometry}
\usepackage{graphicx}
\usepackage{cite}

\newcommand{\mysection}{\setcounter{equation}{0}\section}

\def\beq{\begin{equation}}
\def\eeq{\end{equation}}
\def\beqa{\begin{eqnarray}}
\def\eeqa{\end{eqnarray}}
 
\begin{document}

\begin{center}
{\Large \bf Higher-order corrections in $t{\bar t}\gamma$ cross sections}
\end{center}

\vspace{2mm}

\begin{center}
{\large Nikolaos Kidonakis and Alberto Tonero}\\

\vspace{2mm}

{\it Department of Physics, Kennesaw State University, \\
Kennesaw, GA 30144, USA}

\end{center}

\begin{abstract}
We study higher-order QCD corrections for the associated production of a top-antitop quark pair and a photon ($t{\bar t}\gamma$ production) in proton-proton collisions. We calculate the approximate NNLO cross section, with second-order soft-gluon corrections added to the complete NLO result, including uncertainties from scale dependence and from parton distributions. We compare our results to recent measurements from the LHC, and find that the soft-gluon corrections provide improved agreement with the data. We also calculate differential distributions in top-quark transverse momentum and rapidity.
\end{abstract}

\mysection{Introduction}

The production of a top-antitop quark pair in association with a photon, i.e. $t{\bar t}\gamma$ production, for which evidence was first seen at the Tevatron \cite{CDF}, has been observed and studied at the LHC \cite{ATLAS7,CMS8,ATLAS8,ATLAS13a,ATLAS13b,CMS13a,CMS13b,ATLAS13c}. The cross section for this process is sensitive to the top-quark charge \cite{Baur:2001si,Baur:2004uw} as well as any possible modifications of the top-photon interaction vertex from new physics \cite{Bouzas:2012av,Schulze:2016qas,BessidskaiaBylund:2016jvp}. 

The next-to-leading order (NLO) QCD corrections for this process were calculated in \cite{DMZHGW,DZMHGW} and, including top-quark decays, in \cite{KMMS}. Matching with parton shower was done in \cite{AKZT} and further numerical studies were done in \cite{MPT}, while electroweak corrections were calculated in \cite{DZWSL,PSTZ}. More studies with off-shell effects were performed in \cite{BHKWW1,BHKWW2,BHKWW3}. The top quark charge asymmetry in $t{\bar t}\gamma$ production was studied in \cite{JBMS}.

Soft-gluon resummation \cite{NKGS1,NKGS2,NKsingletop,NK2loop,NKsch,NKtW,NKtt2l,NKtch,NK3loop,FK2020} is very important for top-quark processes since the cross section receives large contributions from soft-gluon emission near partonic threshold because of the large top-quark mass. This is well known for numerous $2 \to 2$ top-quark processes, including top-antitop pair production \cite{NKGS1,NKGS2,NK2loop,NKtt2l,NKttaN3LO,NKdoublediff} and single-top production \cite{NKsingletop,NKsch,NKtW,NKtch,NK3loop}. More recently, the resummation formalism has been extended \cite{FK2020} to $2 \to 3$ processes, in particular $tqH$ production \cite{FK2021}, $tq\gamma$ production \cite{NKNY2022}, and $tqZ$ production \cite{NKNYtqZ}. In all these processes, as well as in $t{\bar t} \gamma$ production, the soft-gluon corrections are dominant and account for the majority of the complete corrections at NLO. 

In this paper, we calculate approximate next-to-next-to-leading order (aNNLO) cross sections and top-quark differential distributions for $t{\bar t}\gamma$ production. The aNNLO results are derived by adding second-order soft-gluon corrections to the complete NLO calculation which includes QCD plus electroweak (EW) corrections. In the next section, we describe the soft-gluon resummation formalism for $t{\bar t}\gamma$ partonic processes. In Section 3, we present numeical results for the total cross section through aNNLO. In Section 4, we present results for the top-quark differential distributions in transverse momentum, $p_T$, and rapidity. We conclude in Section 5.   

\mysection{Resummation for $t{\bar t}\gamma$ production}

We begin with the soft-gluon resummation formalism for $t{\bar t}\gamma$ production. At leading order (LO), the parton-level processes are $a(p_a)+b(p_b) \to t(p_t)+{\bar t}(p_{\bar t})+\gamma(p_{\gamma})$, with $a$ and $b$ denoting two incoming gluons or a quark-antiquark pair, and we define the usual kinematical variables $s=(p_a+p_b)^2$, $t=(p_a-p_t)^2$, and $u=(p_b-p_t)^2$, as well as $s'=(p_t+p_{\bar t})^2$, $t'=(p_b-p_{\bar t})^2$, and $u'=(p_a-p_{\bar t})^2$. If an additional gluon is emitted in the final state, then momentum conservation gives $p_a +p_b=p_t +p_{\bar t} +p_{\gamma}+p_g$ where $p_g$ is the gluon momentum. We also define a threshold variable $s_4=(p_{\bar t}+p_{\gamma}+p_g)^2-(p_{\bar t}+p_{\gamma})^2=s+t+u-m_t^2-(p_{\bar t}+p_{\gamma})^2$ which involves the extra energy from gluon emission and which vanishes as $p_g \to 0$. The soft-gluon corrections appear in the perturbative series as terms with coefficients multiplying plus distributions of logarithms of $s_4$, specifically $\ln^k(s_4/m_t^2)/s_4$ where $k$ takes integer values from 0 through $2n-1$ for the $n$th order corrections.

Resummation follows from factorization properties of the cross section and renormalization-group evolution. We begin by writing the differential cross section for $t{\bar t}\gamma$ production in proton-proton collisions as a convolution, 
\beq
d\sigma_{pp \to t{\bar t}\gamma}=\sum_{a,b} \; 
\int dx_a \, dx_b \,  \phi_{a/p}(x_a, \mu_F) \, \phi_{b/p}(x_b, \mu_F) \, 
d{\hat \sigma}_{ab \to t{\bar t}\gamma}(s_4, \mu_F) \, ,
\label{sigma}
\eeq
where $\mu_F$ is the factorization scale, $\phi_{a/p}$  and $\phi_{b/p}$ are parton distribution functions (pdf) for parton $a$ and parton $b$, respectively, in the proton, and ${\hat \sigma}_{ab \to t{\bar t}\gamma}$ is the partonic cross section which at a given order also depends on the renormalization scale $\mu_R$.

The cross section factorizes if we take Laplace transforms, defined by 
\beq
 {\tilde{\hat\sigma}}_{ab \to t{\bar t}\gamma}(N)=\int_0^s \frac{ds_4}{s} \,  e^{-N s_4/s} \; {\hat\sigma}_{ab \to t{\bar t}\gamma}(s_4), 
\eeq
where $N$ is the transform variable. Under transforms, the logarithms of $s_4$ in the perturbative series turn into logarithms of $N$ which, as we will see, exponentiate. We also define transforms of the pdf via ${\tilde \phi}(N)=\int_0^1 e^{-N(1-x)} \phi(x) \, dx$. Replacing the colliding protons by partons in Eq. (\ref{sigma}) \cite{NKGS2,FK2020,GS}, we thus have a factorized form in transform space
\beq
d{\tilde \sigma}_{ab \to t{\bar t}\gamma}(N)= {\tilde \phi}_{a/a}(N_a, \mu_F) \, {\tilde \phi}_{b/b}(N_b, \mu_F) \, d{\tilde{\hat \sigma}}_{ab \to t{\bar t}\gamma}(N, \mu_F) \, .
\label{factcs}
\eeq

The cross section can be refactorized \cite{NKGS1,NKGS2,FK2020} in terms of an infrared-safe short-distance hard function, $H_{ab \to t{\bar t}\gamma}$,  and a soft function, $S_{ab \to t{\bar t}\gamma}$, which describes the emission of noncollinear soft gluons. The hard and the soft functions are $2\times 2$ matrices in the color space of the partonic scattering for an incoming quark-antiquark pair, while they are $3\times 3$ matrices for two incoming gluons. We have
\beq
d{\tilde{\sigma}}_{ab \to t{\bar t}\gamma}(N)={\tilde \psi}_{a/a}(N_a,\mu_F) \, {\tilde \psi}_{b/b}(N_b,\mu_F) \, {\rm tr} \left\{H_{ab \to t{\bar t}\gamma} \left(\alpha_s(\mu_R)\right) \, {\tilde S}_{ab \to t{\bar t}\gamma} \left(\frac{\sqrt{s}}{N \mu_F} \right)\right\} \, ,
\label{refactcs}
\eeq
where the functions $\psi$ are distributions for incoming partons at fixed value of momentum and involve collinear emission \cite{NKGS1,NKGS2,FK2020,GS}.

Comparing Eqs. (\ref{factcs}) and (\ref{refactcs}), we find an expression for the hard-scattering partonic cross section in transform space
\beq
d{\tilde{\hat \sigma}}_{ab \to t{\bar t}\gamma}(N)=
\frac{{\tilde \psi}_{a/a}(N_a, \mu_F) \, {\tilde \psi}_{b/b}(N_b, \mu_F)}{{\tilde \phi}_{a/a}(N_a, \mu_F) \, {\tilde \phi}_{b/b}(N_b, \mu_F)} \; \,  {\rm tr} \left\{H_{ab \to t{\bar t}\gamma}\left(\alpha_s(\mu_R)\right) \, {\tilde S}_{ab \to t{\bar t}\gamma}\left(\frac{\sqrt{s}}{N \mu_F} \right)\right\} \, .
\label{sigN}
\eeq

The dependence of the soft matrix on the transform variable, $N$, is resummed via renormalization-group evolution \cite{NKGS1,NKGS2}. Thus, ${\tilde S}_{ab \to t{\bar t}\gamma}$ obeys a renormalization-group equation involving a soft anomalous dimension matrix, $\Gamma_{\! S \, ab \to t{\bar t}\gamma}$, which is calculated from the coefficients of the ultraviolet poles of the eikonal diagrams for the partonic processes \cite{NKGS1,NKGS2,NKsingletop,NK2loop,NKsch,NKtW,NKtt2l,NKtch,NK3loop,FK2020}.

The $N$-space resummed cross section, which resums logarithms of $N$, is derived from the renormalization-group evolution of the functions ${\tilde S}_{ab \to t{\bar t}\gamma}$, ${\tilde \psi}_{a/a}$, ${\tilde \psi}_{b/b}$, ${\tilde \phi}_{a/a}$, and ${\tilde \phi}_{b/b}$ in Eq. (\ref{sigN}), and it is given by
\beqa
d{\tilde{\hat \sigma}}_{ab \to t{\bar t}\gamma}^{\rm resum}(N) &=&
\exp\left[\sum_{i=a,b} E_{i}(N_i)\right] \, 
\exp\left[\sum_{i=a,b} 2 \int_{\mu_F}^{\sqrt{s}} \frac{d\mu}{\mu} \gamma_{i/i}(N_i)\right]
\nonumber\\ && \hspace{-5mm} \times \,
{\rm tr} \left\{H_{ab \to t{\bar t}\gamma}\left(\alpha_s(\sqrt{s})\right) {\bar P} \exp \left[\int_{\sqrt{s}}^{{\sqrt{s}}/N}
\frac{d\mu}{\mu} \; \Gamma_{\! S \, ab \to t{\bar t}\gamma}^{\dagger} \left(\alpha_s(\mu)\right)\right] \; \right.
\nonumber\\ && \left. \hspace{5mm} \times \,
{\tilde S}_{ab \to t{\bar t}\gamma} \left(\alpha_s\left(\frac{\sqrt{s}}{N}\right)\right) \;
P \exp \left[\int_{\sqrt{s}}^{{\sqrt{s}}/N}
\frac{d\mu}{\mu}\; \Gamma_{\! S \, ab \to t{\bar t}\gamma}
\left(\alpha_s(\mu)\right)\right] \right\} \, ,
\nonumber \\
\label{resummed}
\eeqa
where $P$ (${\bar P}$) denotes path-ordering in the same (reverse) sense as the integration variable $\mu$.
The first exponential in Eq. (\ref{resummed}) resums soft and collinear emission from the initial-state partons \cite{GS}, while the second exponential involves the parton anomalous dimensions $\gamma_{i/i}$ and the factorization scale $\mu_F$. These are followed by exponentials with the soft anomalous dimension matrix $\Gamma_{\! S \, ab \to t{\bar t}\gamma }$ and its Hermitian adjoint. Explicit results for the first two exponents of Eq. (\ref{resummed}) can be found in Ref. \cite{FK2020}.

The soft anomalous dimensions for $t{\bar t} \gamma$ production are essentially the same as for $t{\bar t}$ production \cite{NKGS1,NKGS2,NKtt2l}, since the color structure of the hard scattering is the same, with only some modifications to account for the $2 \to 3$ kinematics (see Ref. \cite{NKuni} for a review). For the processes  $q(p_a)+{\bar q}(p_b) \to t(p_t)+{\bar t}(p_{\bar t})+\gamma(p_{\gamma})$ we choose a color tensor basis of $s$-channel singlet and octet exchange, $c_1^{q{\bar q}\rightarrow t{\bar t}\gamma} = \delta_{ab}\delta_{12}$,  $c_2^{q{\bar q}\rightarrow t{\bar t}\gamma} =  T^c_{ba} \, T^c_{12}$, where 1 and 2 are color indices for the top and antitop quarks, respectively. Then, the four matrix elements of $\Gamma_{\!\! S \, q{\bar q}\rightarrow t{\bar t}\gamma}$ are given at one loop \cite{NKuni} by 
\beqa
\Gamma_{\!\! 11 \, q{\bar q}\rightarrow t{\bar t}\gamma}^{(1)} &=& -C_F\left(L_{\beta'}+1\right),
\quad
\Gamma_{\!\! 12 \, q{\bar q}\rightarrow t{\bar t}\gamma}^{(1)}=
\frac{C_F}{2N_c} \Gamma_{\!\! 21 \, q{\bar q}\rightarrow t{\bar t}\gamma}^{(1)} ,
\quad
\Gamma_{\!\! 21 \, q{\bar q}\rightarrow t{\bar t}\gamma}^{(1)}=
\ln\left(\frac{(t-m_t^2)(t'-m_t^2)}{(u-m_t^2)(u'-m_t^2)}\right),
\nonumber \\ 
\Gamma_{\!\! 22 \, q{\bar q}\rightarrow t{\bar t}\gamma}^{(1)} &=& \left(C_F-\frac{C_A}{2}\right)
\left[-L_{\beta'}-1+2\ln\left(\frac{(t-m_t^2)(t'-m_t^2)}{(u-m_t^2)(u'-m_t^2)}\right)\right]
+\frac{C_A}{2}\left[\ln\left(\frac{(t-m_t^2)(t'-m_t^2)}{s\, m_t^2}\right)-1\right]
\nonumber \\
\eeqa
where $L_{\beta'}=\frac{(1+{\beta'}^2)}{2\beta'}\ln\left(\frac{1-\beta'}{1+\beta'}\right)$, with $\beta'=\sqrt{1-4m_t^2/s'}$, and $C_F=4/3$, $C_A=N_c=3$ in QCD. 

For the processes $g(p_a)+g(p_b) \to t(p_t)+{\bar t}(p_{\bar t})+\gamma(p_{\gamma})$ in the color tensor basis 
$c_1^{gg\rightarrow t{\bar t}\gamma}=\delta^{ab}\,\delta_{12}$, $c_2^{gg\rightarrow t{\bar t}\gamma}=d^{abc}\,T^c_{12}$, $c_3^{gg\rightarrow t{\bar t}\gamma}=i f^{abc}\,T^c_{12}$, the nine matrix elements of $\Gamma_{\!\! S \, gg\rightarrow t{\bar t}\gamma}$ are given at one loop \cite{NKuni} by 
\beqa
\Gamma_{\!\! 11 \, gg\rightarrow t{\bar t}\gamma}^{(1)}&=& -C_F\left(L_{\beta'}+1\right)  \, , \quad
\Gamma_{\!\! 12 \, gg\rightarrow t{\bar t}\gamma}^{(1)}=\Gamma_{\!\! 21 \, gg\rightarrow t{\bar t}\gamma}^{(1)}=0,
\nonumber \\
\Gamma_{\!\! 13 \, gg\rightarrow t{\bar t}\gamma}^{(1)}&=& \frac{1}{2}\ln\left(\frac{(t-m_t^2)(t'-m_t^2)}{(u-m_t^2)(u'-m_t^2)}\right) \, , 
\quad \Gamma_{\!\! 31 \, gg\rightarrow t{\bar t}\gamma}^{(1)} = 2 \, \Gamma_{\!\! 13 \, gg\rightarrow t{\bar t}\gamma}^{(1)} \, ,
\nonumber \\
\Gamma_{\!\! 22 \, gg\rightarrow t{\bar t}\gamma}^{(1)}&=& \left(C_F-\frac{C_A}{2}\right) \left(-L_{\beta'}-1\right)
+\frac{C_A}{2}\left[\frac{1}{2}\ln\left(\frac{(t-m_t^2)(t'-m_t^2)(u-m_t^2)(u'-m_t^2)}{s^2\, m_t^4}\right)-1\right] \, ,
\nonumber \\
\Gamma_{\!\! 23 \, gg\rightarrow t{\bar t}\gamma}^{(1)}&=&\frac{C_A}{2} \, \Gamma_{\!\! 13 \, gg\rightarrow t{\bar t}\gamma}^{(1)} \, , \quad 
\Gamma_{\!\! 32 \, gg\rightarrow t{\bar t}\gamma}^{(1)}=\frac{(N_c^2-4)}{2N_c} \, \Gamma_{\!\! 13 \, gg\rightarrow t{\bar t}\gamma}^{(1)} \, , \quad
\Gamma_{\!\! 33 \, gg\rightarrow t{\bar t}\gamma}^{(1)}=\Gamma_{\!\! 22 \, gg\rightarrow t{\bar t}\gamma}^{(1)} \, .
\eeqa

At two loops, again the results are essentially the same as for $t{\bar t}$ production \cite{NKtt2l,NKuni}. 

We can expand the resummed cross section, Eq. (\ref{resummed}), to any fixed order, and then do a straightforward inversion back to momentum space without requiring a prescription \cite{FK2020}. Thus, by expanding to NNLO, we can calculate the first-order and second-order soft-gluon corrections. By adding the latter to the complete NLO results, we thus perform aNNLO calculations. 

\mysection{Total cross sections}

In this section we present results for total cross sections for $t{\bar t}\gamma$ production. We use a top-quark mass $m_t=172.5$ GeV and set the factorization and renormalization scales equal to each other, with this common scale denoted by $\mu$. The complete NLO results, which include both QCD and EW corrections, are calculated using {\small \sc MadGraph5\_aMC@NLO} \cite{MG5} following the prescriptions of Ref. \cite{PSTZ}. We use MSHT20 \cite{MSHT20} pdf in our calculations.

\begin{figure}[htbp]
\begin{center}
\includegraphics[width=88mm]{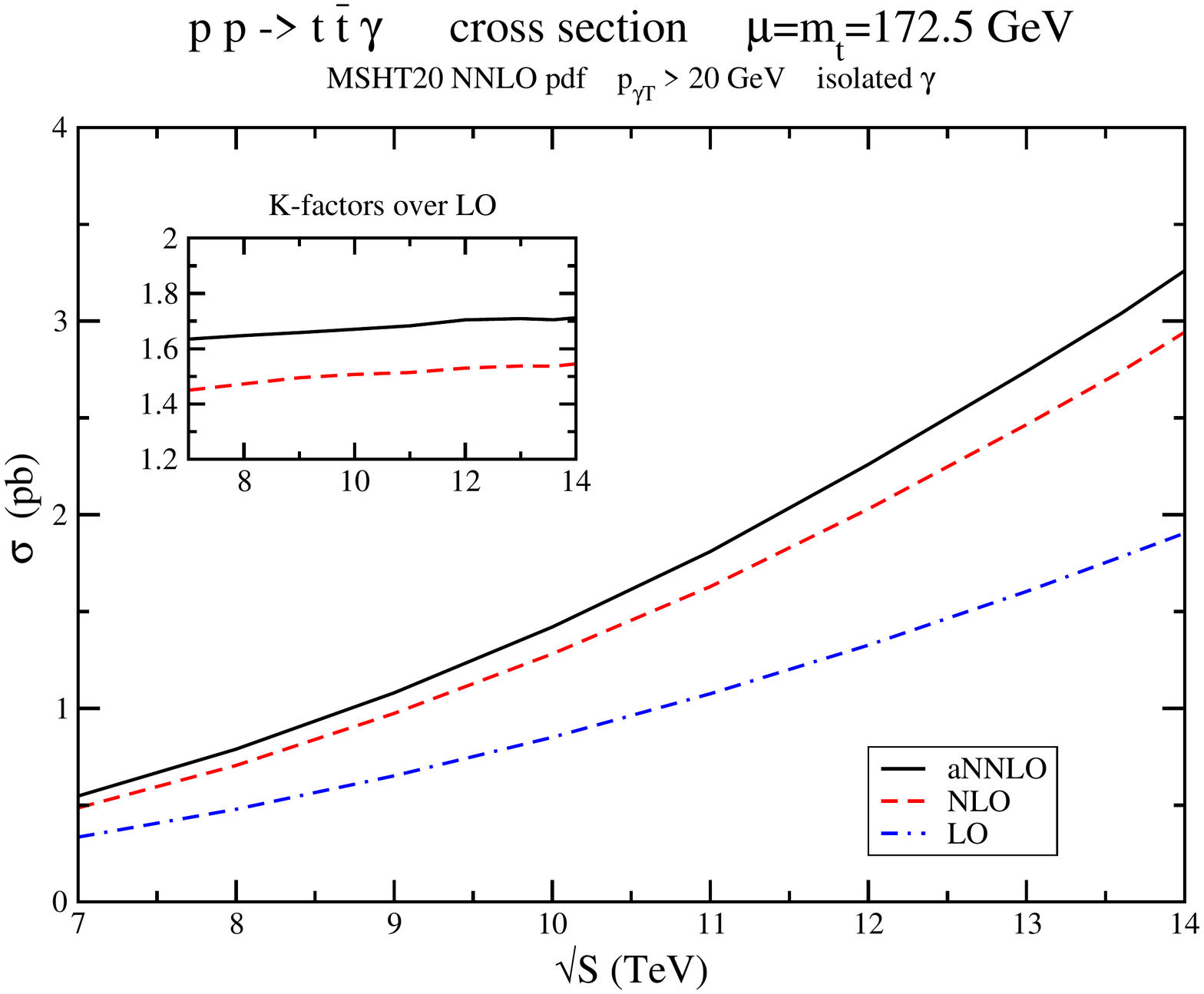}
\includegraphics[width=88mm]{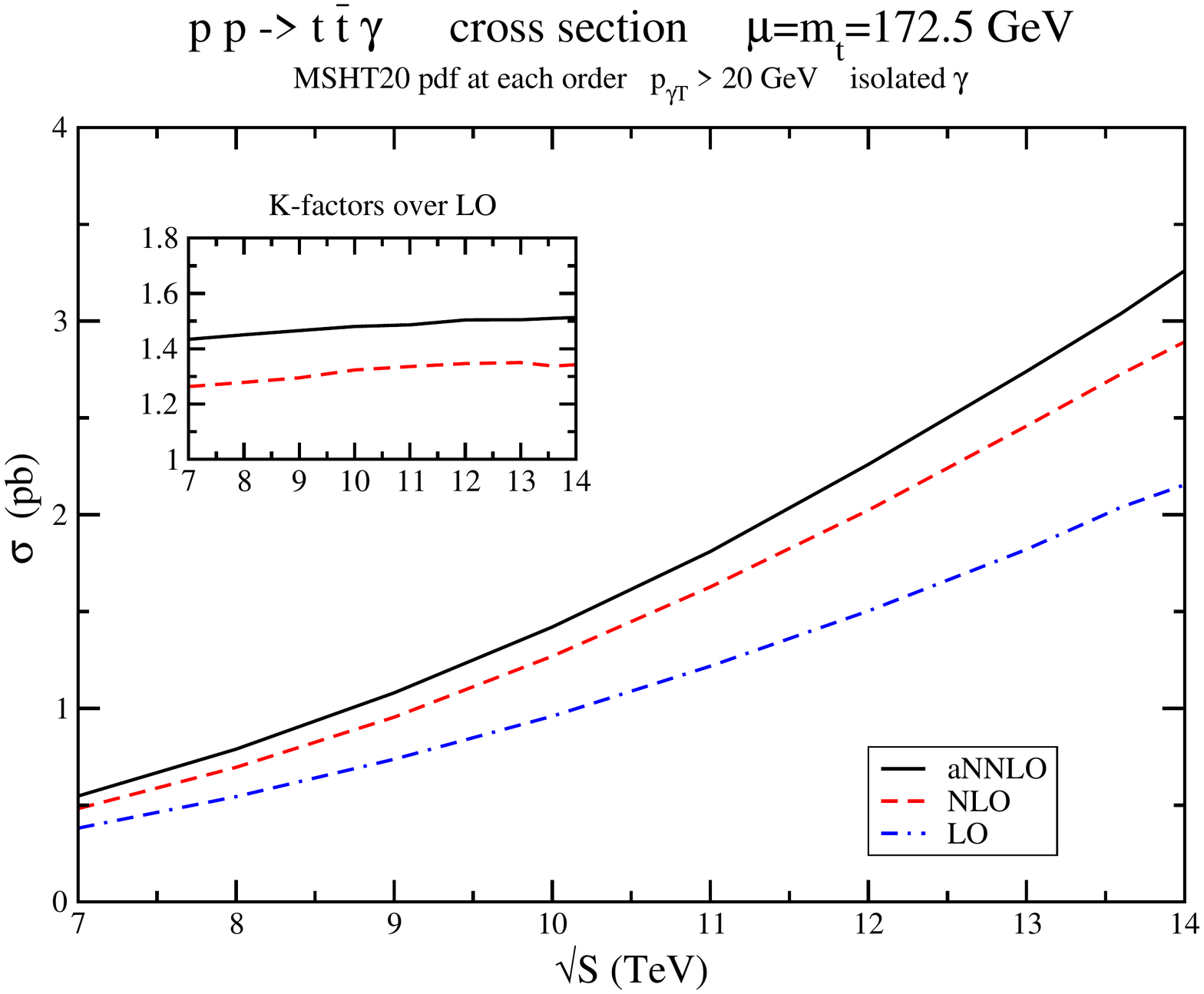}
\caption{The total cross sections at LO, NLO, and aNNLO for $t{\bar t}\gamma$ production in $pp$ collisions at LHC energies using (left) MSHT20 NNLO pdf and (right) MSHT20 pdf at each order. The inset plots display the NLO/LO and aNNLO/LO $K$-factors.}
\label{xsec}
\end{center}
\end{figure}

In Fig. \ref{xsec}, we plot the total cross section for  $t{\bar t}\gamma$ production in proton-proton collisions as a function of the collider energy. We impose a cut on the transverse momentum of the photon, $p_{\gamma T}>20$ GeV, and require the photon to be isolated, as defined in Ref. \cite{KMMS}. We show the central ($\mu=m_t$) LO, NLO, and aNNLO cross sections using MSHT20 NNLO pdf throughout (left plot) or using MSHT20 pdf corresponding to each order of the perturbative calculation (right plot).  Thus, the plot on the left is more useful in showing how each order in the series contributes to the final aNNLO result, while the plot on the right is more useful in showing stand-alone results at each order.
The inset plots in both cases display the $K$-factors relative to LO, i.e. the NLO/LO and aNNLO/LO ratios. The $K$-factors are large and they rise slowly with collision energy.

\begin{table}[htbp]
\begin{center}
\begin{tabular}{|c|c|c|c|c|c|c|c|c|} \hline
\multicolumn{6}{|c|}{$t{\bar t} \gamma$ cross sections in $pp$ collisions at the LHC, $p_{\gamma \, T} > 20$ GeV, isolated $\gamma$} \\ \hline
$\sigma$ in pb & 7 TeV & 8 TeV & 13 TeV & 13.6 TeV & 14 TeV \\ \hline
LO QCD (LO pdf)  & $0.380^{+0.143}_{-0.096} $ & $0.541^{+0.200}_{-0.135}$ & $1.81^{+0.61}_{-0.42}$ & $2.03^{+0.68}_{-0.47}$ & $2.14^{+0.71}_{-0.50}$ \\ \hline
LO QCD+EW (LO pdf)  & $0.381^{+0.144}_{-0.095} $ & $0.544^{+0.200}_{-0.134}$ & $1.82^{+0.61}_{-0.42}$ & $2.04^{+0.68}_{-0.47}$ & $2.15^{+0.72}_{-0.49}$ \\ \hline
LO QCD (NLO pdf)  & $0.331^{+0.116}_{-0.079} $ & $0.474^{+0.162}_{-0.112}$ & $1.58^{+0.49}_{-0.34}$ & $1.76^{+0.55}_{-0.38}$ & $1.89^{+0.57}_{-0.41}$ \\ \hline
LO QCD+EW (NLO pdf)  & $0.333^{+0.116}_{-0.079} $ & $0.477^{+0.162}_{-0.111}$ & $1.60^{+0.49}_{-0.35}$ & $1.78^{+0.53}_{-0.39}$ & $1.90^{+0.57}_{-0.41}$ \\ \hline
LO QCD (NNLO pdf)  & $0.333^{+0.116}_{-0.080} $ & $0.478^{+0.163}_{-0.113}$ & $1.59^{+0.50}_{-0.35}$ & $1.77^{+0.54}_{-0.39}$ & $1.89^{+0.57}_{-0.41}$ \\ \hline
LO QCD+EW (NNLO pdf)  & $0.335^{+0.116}_{-0.080} $ & $0.479^{+0.162}_{-0.112}$ & $1.60^{+0.49}_{-0.34}$ & $1.78^{+0.54}_{-0.38}$ & $1.90^{+0.58}_{-0.40}$ \\ \hline
NLO QCD (NLO pdf) & $0.491^{+0.064}_{-0.066} $ & $0.707^{+0.092}_{-0.094}$ & $2.47^{+0.35}_{-0.32}$ & $2.72^{+0.36}_{-0.35}$ & $2.92^{+0.38}_{-0.37}$ \\ \hline
NLO QCD+EW (NLO pdf) & $0.482^{+0.058}_{-0.063} $ & $0.695^{+0.085}_{-0.089}$ & $2.46^{+0.33}_{-0.32}$ & $2.73^{+0.36}_{-0.35}$ & $2.89^{+0.37}_{-0.36}$ \\ \hline
NLO QCD (NNLO pdf) & $0.490^{+0.063}_{-0.065} $ & $0.708^{+0.090}_{-0.094}$ & $2.49^{+0.34}_{-0.33}$ & $2.76^{+0.38}_{-0.36}$ & $2.96^{+0.41}_{-0.38}$ \\ \hline
NLO QCD+EW (NNLO pdf) & $0.485^{+0.062}_{-0.063} $ & $0.705^{+0.089}_{-0.092}$ & $2.47^{+0.32}_{-0.32}$ & $2.74^{+0.37}_{-0.35}$ & $2.94^{+0.39}_{-0.37}$ \\ \hline
aNNLO (NNLO pdf) & $0.547^{+0.032}_{-0.027} $ & $0.789^{+0.044}_{-0.040}$ & $2.74^{+0.18}_{-0.16}$ & $3.04^{+0.20}_{-0.16}$ & $3.26^{+0.21}_{-0.17}$ \\ \hline
\end{tabular}
\caption[]{The $t{\bar t}\gamma$ cross sections (in pb) at LO, NLO, and aNNLO, with scale uncertainties, in $pp$ collisions with $\sqrt{S}=7$, 8, 13, 13.6, and 14 TeV, $m_t=172.5$ GeV, and MSHT20 pdf.}
\label{table1}
\end{center}
\end{table}

In Table 1 we show total rates, with scale uncertainties, for $t{\bar t}\gamma$ production for various LHC energies at LO, NLO, and aNNLO using the photon cuts described before and MSHT20 pdf at various orders. For the LO and NLO calculations, we show results both with QCD only contributions and with QCD+EW contributions. We note that the difference between the NLO QCD and NLO QCD+EW cross sections is smaller than 1\%, making the EW effects negligible, in agreement with the results of \cite{PSTZ}.

The central results in Table 1 are obtained by setting $\mu=m_t$ while the uncertainties are found by varying the scales between $m_t/2$ and $2m_t$. The scale uncertainties are roughly $\pm 13$\% at NLO and around 5 to 6\% at aNNLO for all LHC energies. In addition to scale variation, there are also uncertainties from the pdf. Including pdf uncertainties, we find that the aNNLO cross sections are given by $0.547{}^{+0.032+0.011}_{-0.027-0.008}$ pb at 7 TeV;  $0.789^{+0.044+0.016}_{-0.040-0.010}$ pb at 8 TeV; $2.74^{+0.18+0.04}_{-0.16-0.03}$ pb at 13 TeV; $3.04^{+0.20+0.06}_{-0.16-0.03}$ pb at 13.6 TeV; and $3.26^{+0.21+0.06}_{-0.17-0.03}$ pb at 14 TeV. The pdf uncertainties are much smaller than the scale uncertainties, especially at higher energies.

Furthermore, we note that the $K$-factors as well as the scale and pdf uncertainties are very similar to those found in $t{\bar t}$ production \cite{NKtt2l,NKttaN3LO}. While this finding is not too surprising, given the similar QCD structure for these processes, it is nevertheless noteworthy.

Finally, we compare our predictions with the most recent, and most accurate, measurement of the cross section for the process in the dilepton decay channel from CMS at 13 TeV energy \cite{CMS13b}. They measure a cross section of $175.2 \pm 2.5$(stat) $\pm 6.3$(syst) fb which is compared to an NLO prediction of $155 \pm 27$ fb with scale and pdf uncertainties. The CMS central result is significantly higher than the central NLO prediction, and it also has a much smaller uncertainty. Our aNNLO calculation instead gives a cross section of $173^{+11}_{-10}{}^{+3}_{-2}$ fb which is in much better agreement with the data and more commensurate with it in overall uncertainty. 

\mysection{Top-quark differential distributions}

As the resummation formalism in Section 2 indicates, soft-gluon resummation is derived not only for total cross sections but also for differential distributions. In this section we present results for the top-quark transverse-momentum and rapidity distributions in $t{\bar t}\gamma$ production. Again, we use {\small \sc MadGraph5\_aMC@NLO} \cite{MG5} for the complete NLO results to which we add second-order soft-gluon corrections to derive aNNLO distributions.

\begin{figure}[htbp]
\begin{center}
\includegraphics[width=88mm]{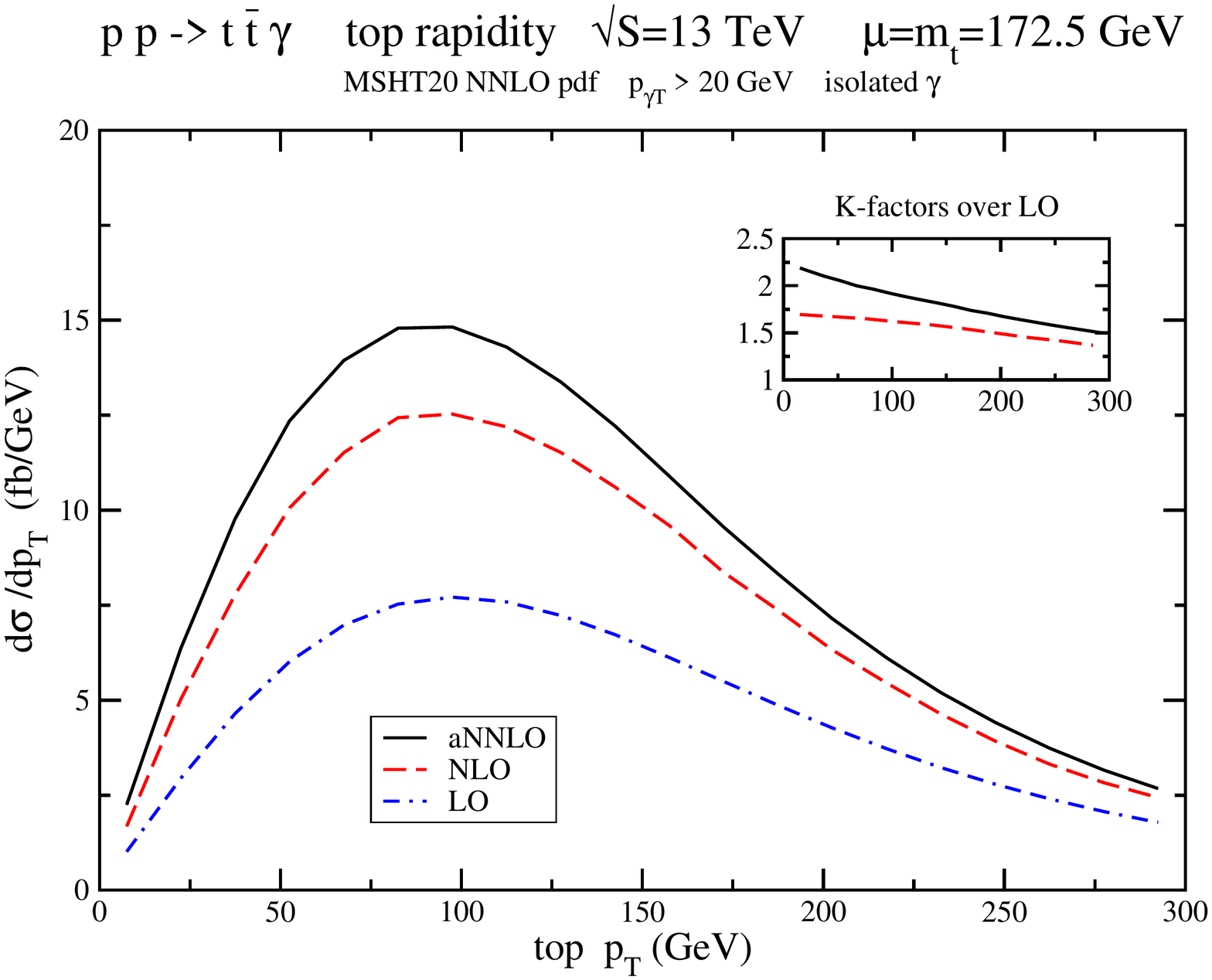}
\includegraphics[width=88mm]{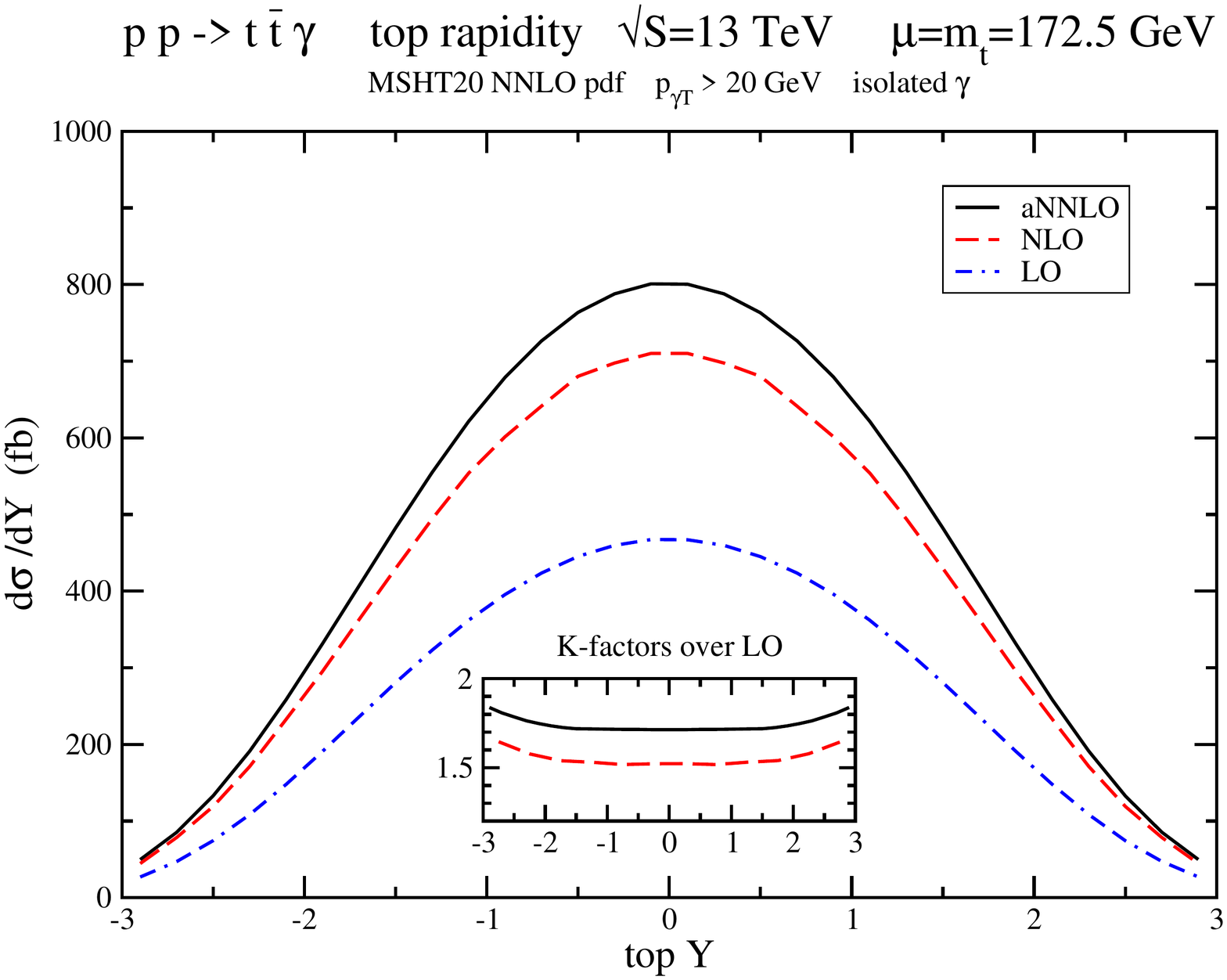}
\caption{Top-quark $p_T$ (left) and rapidity (right) distributions at LO, NLO, and aNNLO in $t{\bar t}\gamma$ production at 13 TeV LHC energy. The inset plots show the NLO/LO and aNNLO/LO $K$-factors.}
\label{ptyt13}
\end{center}
\end{figure}

In Fig. \ref{ptyt13}, we plot the central ($\mu=m_t$) results at LO, NLO, and aNNLO for the top-quark $p_T$ (left plot) and rapidity (right plot) distributions in $t{\bar t}\gamma$ production at 13 TeV LHC energy with MSHT20 NNLO pdf and with the same $p_{\gamma \, T}$ cut and photon isolation requirement as for the total cross section. The $K$-factors relative to the LO results are shown in the inset plots. For the $p_T$ distribution, we note that both the NLO/LO and the aNNLO/LO $K$-factors decrease as the top-quark transverse momentum increases. For the rapidity distribution, we see that both $K$-factors are relatively flat for central and moderate values of the top-quark rapidity, but they begin to grow for rapidities larger than 2. Regarding the theoretical uncertainties from scale variation (which are not  shown in the plots) we note that they are very similar to those for the total cross section in most of the range plotted for the $p_T$ and rapidity distributions, but they become a little smaller at large $p_T$ and bigger at large rapidities. 

\begin{figure}[htbp]
\begin{center}
\includegraphics[width=88mm]{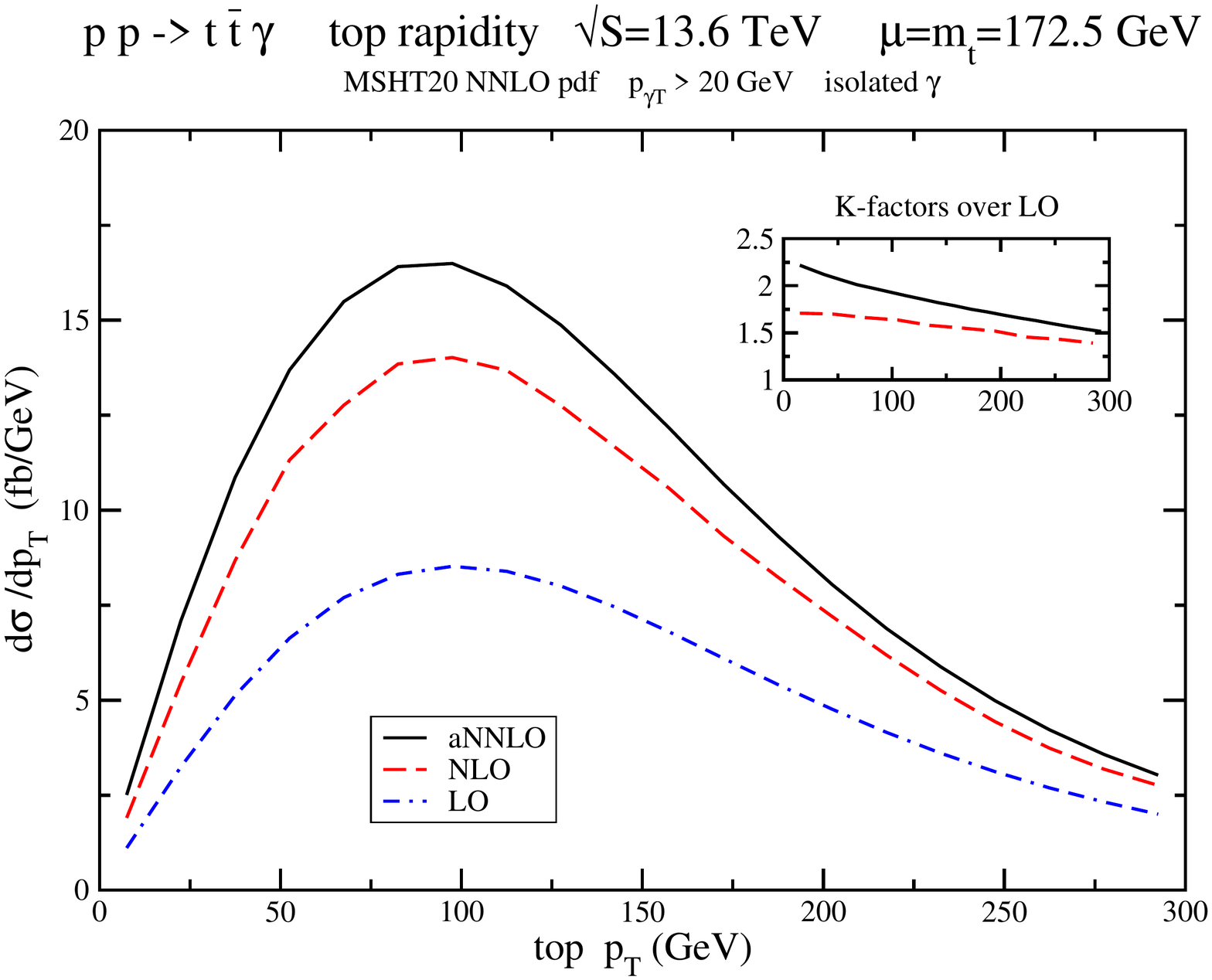}
\includegraphics[width=88mm]{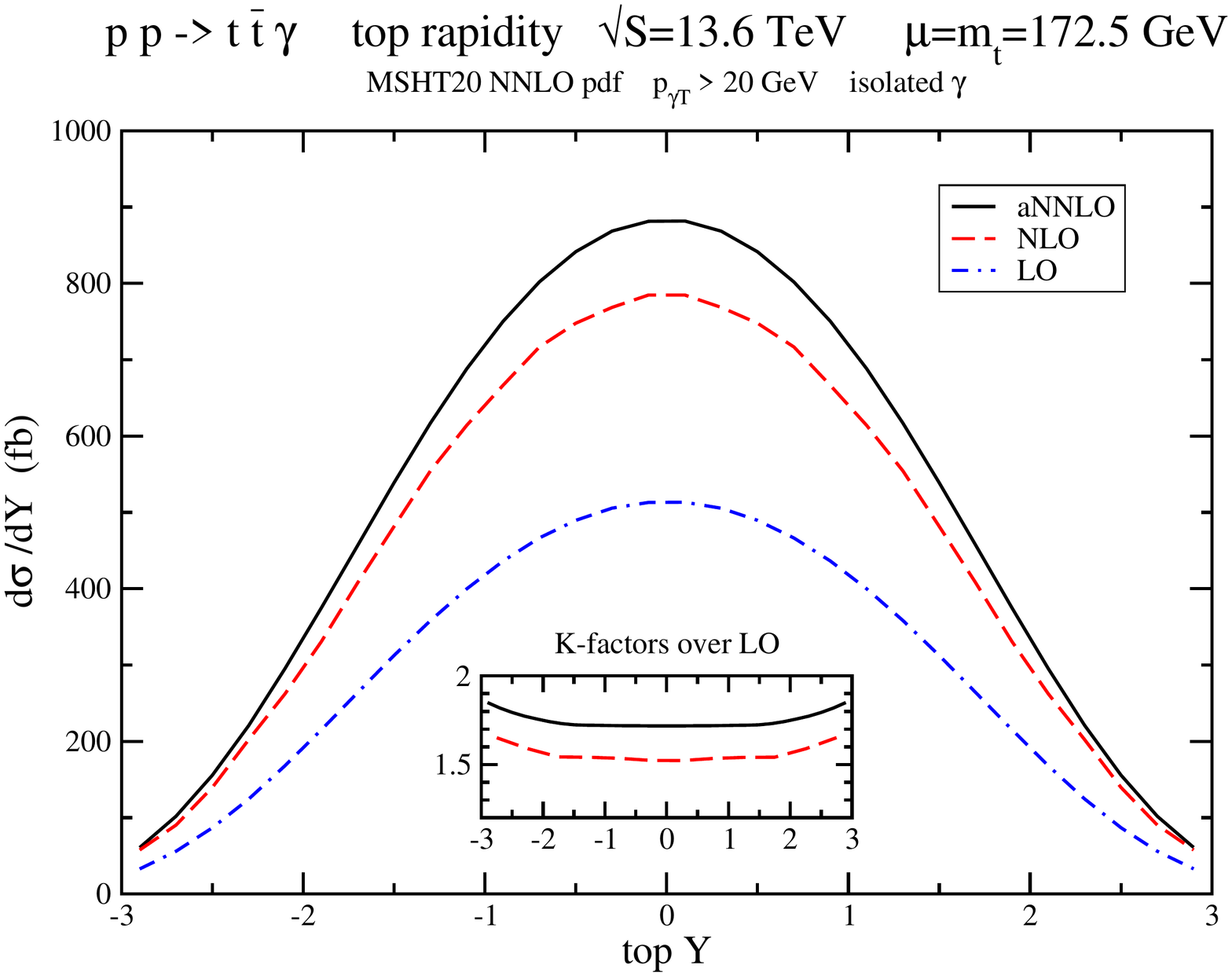}
\caption{Top-quark $p_T$ (left) and rapidity (right) distributions at LO, NLO, and aNNLO in $t{\bar t}\gamma$ production at 13.6 TeV LHC energy. The inset plots show the NLO/LO and aNNLO/LO $K$-factors.}
\label{ptyt13.6}
\end{center}
\end{figure}

In Fig. \ref{ptyt13.6}, we plot the LO, NLO, and aNNLO top-quark $p_T$ (left plot) and rapidity (right plot) distributions in $t{\bar t}\gamma$ production at 13.6 TeV LHC energy. The $K$-factors relative to the LO results are shown in the inset plots, and we observe the same trends for the $K$-factors and the scale uncertainties as at 13 TeV.

\begin{figure}[htbp]
\begin{center}
\includegraphics[width=88mm]{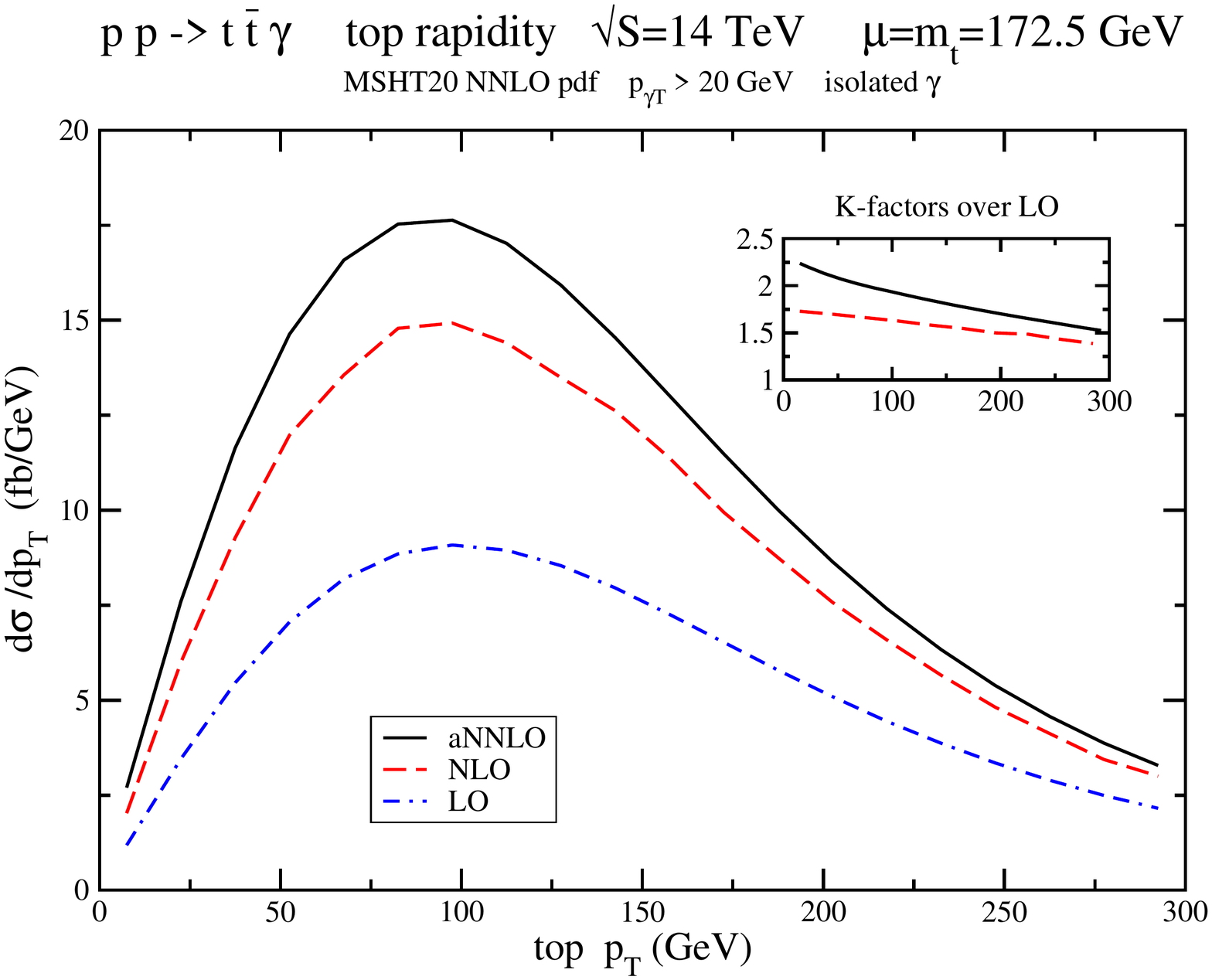}
\includegraphics[width=88mm]{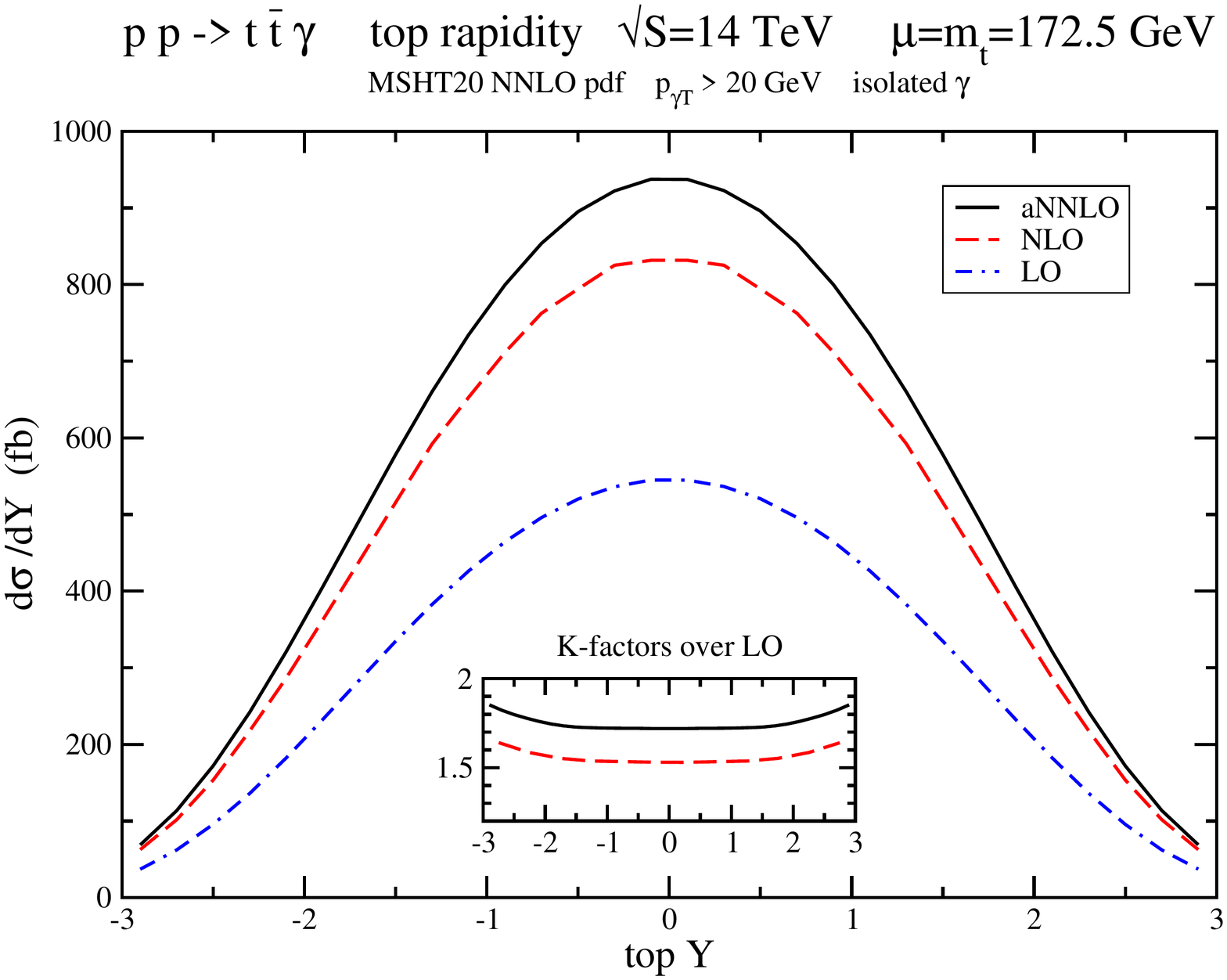}
\caption{Top-quark $p_T$ (left) and rapidity (right) distributions at LO, NLO, and aNNLO in $t{\bar t}\gamma$ production at 14 TeV LHC energy. The inset plots show the NLO/LO and aNNLO/LO $K$-factors.}
\label{ptyt14}
\end{center}
\end{figure}

In Fig. \ref{ptyt14}, we plot the LO, NLO, and aNNLO top-quark $p_T$ (left plot) and rapidity (right plot) distributions in $t{\bar t}\gamma$ production at 14 TeV LHC energy. The $K$-factors relative to the LO results are shown in the inset plots, and again we make the same observations for the $K$-factors and the scale uncertainties as at 13 and 13.6 TeV. In all cases, the enhancement from the aNNLO soft-gluon corrections is very significant.

\mysection{Conclusions}

We have provided a study of higher-order corrections for $t{\bar t} \gamma$ production at the LHC. We have included complete QCD and EW corrections at NLO as well as soft-gluon corrections at aNNLO. We have calculated total cross sections at the LHC as a function of the collider energy, from 7 to 14 TeV, as well as top-quark differential distributions.

The total cross sections get large enhancements from the complete NLO QCD and EW corrections (around 50\%), with the EW effects being negligible, in agreement with the results of \cite{PSTZ}. The additional aNNLO QCD corrections are significant, accounting for a further $\sim 15$\% enhancement of the cross section. The theoretical uncertainties from scale variation get reduced with each increasing perturbative order. We have found that the $K$-factors at both NLO and aNNLO are very similar to the corresponding ones for $t{\bar t}$ production.

We have also computed the top-quark $p_T$ and rapidity distributions for three representative center-of-mass energies, namely 13, 13.6, and 14 TeV. These differential distributions also receive large corrections at NLO and further significant enhancements at aNNLO.

Our aNNLO results provide the most accurate theoretical prediction for $t{\bar t} \gamma$ production at the LHC. Indeed, the enhancement from the aNNLO soft-gluon corrections improves the agreement with recent data from the LHC.

\section*{Acknowledgements}
This material is based upon work supported by the National Science Foundation under Grant No. PHY 2112025.


\begin{thebibliography}{99}

\bibitem{CDF}
CDF collaboration, ``Evidence for $t{\bar t}\gamma$ production and measurement of $\sigma_{t{\bar t} \gamma} / \sigma_{t{\bar t}}$,'' Phys. Rev. D {\bf 84}, 031104 (2011) [arXiv:1106.3970].

\bibitem{ATLAS7}
ATLAS Collaboration, ``Observation of top-quark pair production in association with a photon and measurement of the $t{\bar t}\gamma$ production cross section in $pp$ collisions at ${\sqrt s}= 7$ TeV using the ATLAS detector,'' Phys. Rev. D {\bf 91}, 072007 (2015) [arXiv:1502.00586].

\bibitem{CMS8}
CMS Collaboration, ``Measurement of the semileptonic $t{\bar t}+\gamma$ production cross section in $pp$ collisions at ${\sqrt s} = 8$ TeV,'' JHEP {\bf 10}, 006 (2017) [arXiv:1706.08128]. 

\bibitem{ATLAS8}
ATLAS Collaboration, ``Measurement of the $t{\bar t}\gamma$ production cross section in proton-proton collisions at ${\sqrt s} = 8$ TeV with the ATLAS detector,'' JHEP {\bf 11}, 086 (2017) [arXiv:1706.03046].   

\bibitem{ATLAS13a}
ATLAS Collaboration, ``Measurements of inclusive and differential fiducial cross-sections of $t{\bar t}\gamma$ production in leptonic final states at ${\sqrt s} = 13$ TeV in ATLAS,'' Eur. Phys. J. C {\bf 79}, 382 (2019) [arXiv:1812.01697].

\bibitem{ATLAS13b}
ATLAS Collaboration, ``Measurements of inclusive and differential
cross-sections of combined $t{\bar t}\gamma$ and $tW\gamma$ production
in the $e\mu$ channel at 13 TeV with the ATLAS detector,'' JHEP {\bf 09}, 049 (2020) [arXiv:2007.06946]. 

\bibitem{CMS13a}
CMS Collaboration, ``Measurement of the inclusive and differential $t{\bar t}\gamma$ cross sections in the single-lepton channel and EFT interpretation at ${\sqrt s} = 13$ TeV,'' JHEP {\bf 12}, 180 (2021) [arXiv:2107.01508].

\bibitem{CMS13b}
CMS Collaboration, ``Measurement of the inclusive and differential $t{\bar t}\gamma$ cross sections in the dilepton channel and effective field theory interpretation in proton-proton collisions at ${\sqrt s} = 13$ TeV,''  JHEP {\bf 05}, 091 (2022) [arXiv:2201.07301].  

\bibitem{ATLAS13c}
ATLAS Collaboration, ``Measurement of the charge asymmetry in top quark pair production in association with a photon with the ATLAS experiment,'' ATLAS-CONF-2022-049.

\bibitem{Baur:2001si}
U. Baur, M. Buice, and L.H. Orr, ``Direct measurement of the top quark charge at hadron colliders,''Phys. Rev. D {\bf 64}, 094019 (2001) [arXiv:hep-ph/0106341].

\bibitem{Baur:2004uw}
U. Baur, A. Juste, L.H. Orr, and D. Rainwater, ``Probing electroweak top quark couplings at hadron colliders,'' Phys. Rev. D {\bf 71}, 054013 (2005) [arXiv:hep-ph/0412021].
 
\bibitem{Bouzas:2012av}
A.O. Bouzas and F. Larios, ``Electromagnetic dipole moments of the Top quark,'' Phys. Rev. D {\bf 87}, 074015 (2013) [arXiv:1212.6575].
 
\bibitem{BessidskaiaBylund:2016jvp}
O. Bessidskaia Bylund, F. Maltoni, I. Tsinikos, E. Vryonidou, and C. Zhang, ``Probing top quark neutral couplings in the Standard Model Effective Field Theory at NLO in QCD,'' JHEP {\bf 05}, 052 (2016) [arXiv:1601.08193].

\bibitem{Schulze:2016qas}
M. Schulze and Y. Soreq, ``Pinning down electroweak dipole operators of the top quark,'' Eur. Phys. J. C {\bf 76}, 466 (2016) [arXiv:1603.08911].

\bibitem{DMZHGW}
P.-F. Duan, W.-G. Ma, R.-Y. Zhang, L. Han, L. Guo, and S.-M. Wang, ``QCD corrections to associated production of $t {\bar t}\gamma$ at hadron colliders, Phys. Rev. D {\bf 80}, 014022 (2009) [arXiv:0907.1324].

\bibitem{DZMHGW}
P.-F. Duan, R.-Y. Zhang, W.-G. Ma, L. Han, L. Guo, and S.-M. Wang, ``Next-to-leading order QCD corrections to  $t {\bar t}\gamma$ production at the 7 TeV LHC,'' Chin. Phys. Lett. {\bf 28}, 111401 (2011) [arXiv:1110.2315].

\bibitem{KMMS}
K. Melnikov, M. Schulze, and A. Scharf, ``QCD corrections to top quark pair production in association with a photon at hadron colliders,'' Phys. Rev. D {\bf 83}, 074013 (2011) [arXiv:1102.1967].      

\bibitem{AKZT}
A. Kardos and Z. Trocsanyi, ``Hadroproduction of t anti-t pair in association with an isolated photon at NLO accuracy matched with parton shower,'' JHEP {\bf 05}, 090 (2015) [arXiv:1406.2324].

\bibitem{MPT}
F. Maltoni, D. Pagani, and I. Tsinikos, `` Associated production of a top-quark pair with vector bosons at NLO in QCD: impact on $t{\bar t}H$ searches at the LHC,'' JHEP {\bf 02}, 113 (2016) [arXiv:1507.05640].    

\bibitem{DZWSL}
P.-F. Duan, Y. Zhang, Y. Wang, M. Song, and G. Li, ``Electroweak corrections to top quark pair production in association with a hard photon at hadron colliders, Phys. Lett. B {\bf 766}, 102 (2017) [arXiv:1612.00248].

\bibitem{PSTZ}
D. Pagani, H.-S. Shao, I. Tsinikos, and M. Zaro, ``Automated EW corrections with isolated photons: $t{\bar t}\gamma$, $t{\bar t}\gamma\gamma$ and $t\gamma j$ as case studies,'' JHEP {\bf 09}, 155 (2021) [arXiv:2106.02059].

\bibitem{BHKWW1}
G. Bevilacqua, H.B. Hartanto, M. Kraus, T. Weber, and M. Worek, ``Hard photons in hadroproduction of top quarks with realistic final states,'' JHEP {\bf 10}, 158 (2018) [arXiv:1803.09916].

\bibitem{BHKWW2}
G. Bevilacqua, H.B. Hartanto, M. Kraus, T. Weber, and M. Worek, ``Precise predictions for $t{\bar t} \gamma / t{\bar t}$ cross section ratios at the LHC,'' JHEP {\bf 01}, 188 (2019) [arXiv:1809.08562].

\bibitem{BHKWW3}
G. Bevilacqua, H. B. Hartanto, M. Kraus, T. Weber, and M. Worek, ``Off-shell vs on-shell modelling of top quarks in photon associated production,'' JHEP {\bf 03}, 154 (2020) [arXiv:1912.09999].

\bibitem{JBMS}
J. Bergner and M. Schulze, ``The top quark charge asymmetry in $t{\bar t}\gamma$ production at the LHC,'' Eur. Phys. J. C {\bf 79}, 189 (2019) [arXiv:1812.10535].

\bibitem{NKGS1}
N. Kidonakis and G. Sterman, ``Subleading logarithms in QCD hard scattering,'' Phys. Lett. B {\bf 387}, 867 (1996). 

\bibitem{NKGS2}
N. Kidonakis and G. Sterman, ``Resummation for QCD hard scattering,'' Nucl. Phys. B {\bf 505}, 321 (1997) [hep-ph/9705234].

\bibitem{NKsingletop}
N. Kidonakis, ``Single top production at the Tevatron: threshold resummation and finite-order soft gluon corrections,'' Phys. Rev. D {\bf 74}, 114012 (2006) [arXiv:hep-ph/0609287]. 

\bibitem{NK2loop}
N. Kidonakis, ``Two-loop soft anomalous dimensions and NNLL resummation for heavy quark production,'' Phys. Rev. Lett. {\bf 102}, 232003 (2009) [arXiv:0903.2561].

\bibitem{NKsch}
N. Kidonakis, ``NNLL resummation for $s$-channel single top quark production,'' Phys. Rev. D {\bf 81}, 054028 (2010) [arXiv:1001.5034].

\bibitem{NKtW}
N. Kidonakis, ``Two-loop soft anomalous dimensions for single top quark associated production with a $W^-$ or $H^-$,'' Phys. Rev. D {\bf 82}, 054018 (2010) [arXiv:1005.4451]. 

\bibitem{NKtt2l}
N. Kidonakis, ``Next-to-next-to-leading soft-gluon corrections for the top quark cross section and transverse momentum distribution,'' Phys. Rev. D {\bf 82}, 114030 (2010) [arXiv:1009.4935].

\bibitem{NKtch}
N. Kidonakis, ``Next-to-next-to-leading-order collinear and soft gluon corrections for $t$-channel single top quark production,'' Phys. Rev. D {\bf 83}, 091503 (2011) [arXiv:1103.2792]. 

\bibitem{NK3loop}
N. Kidonakis, ``Soft anomalous dimensions for single-top production at three loops,'' Phys. Rev. D {\bf 99}, 074024 (2019) [arXiv:1901.09928].

\bibitem{FK2020}
M. Forslund and N. Kidonakis, ``Resummation for $2 \to n$ processes in single-particle-inclusive kinematics,'' Phys. Rev. D {\bf 102}, 034006 (2020) [arXiv:2003.09021].

\bibitem{NKttaN3LO}
N. Kidonakis, ``NNNLO soft-gluon corrections for the top-antitop pair production cross section,'' Phys. Rev. D {\bf 90}, 014006 (2014) [arXiv:1405.7046].

\bibitem{NKdoublediff}
N. Kidonakis, ``Top-quark double-differential distributions at approximate N$^3$LO,'' Phys. Rev. D {\bf 101}, 074006 (2020) [arXiv:1912.10362].

\bibitem{FK2021}
M. Forslund and N. Kidonakis, ``Soft-gluon corrections for the associated production of a single top quark and a Higgs boson,'' Phys. Rev. D {\bf 104}, 034024 (2021) [arXiv:2103.01228].

\bibitem{NKNY2022}
N. Kidonakis and N. Yamanaka, ``QCD corrections in $tq\gamma$ production at hadron colliders,'' Eur. Phys. J. C {\bf 82}, 670 (2022) [arXiv:2201.12877].

\bibitem{NKNYtqZ}
N. Kidonakis and N. Yamanaka, ``Soft-gluon corrections for $tqZ$ production,'' arXiv:2210.09542.

\bibitem{GS}
G. Sterman, ``Summation of large corrections to short-distance hadronic cross sections,'' Nucl. Phys. B {\bf 281}, 310 (1987). 

\bibitem{NKuni}
N. Kidonakis, ``Soft anomalous dimensions and resummation in QCD,'' Universe {\bf 6}, 165 (2020) [arXiv:2008.09914].

\bibitem{MG5}
J. Alwall {\it et al.}, ``The automated computation of tree-level and next-to-leading order differential cross sections, and their matching to parton shower simulations,'' JHEP {\bf 07}, 079 (2014) [arXiv:1405.0301].

\bibitem{MSHT20}
S. Bailey, T. Cridge, L.A. Harland-Lang, A.D. Martin, and R.S. Thorne, ``Parton distributions from LHC, HERA, Tevatron and fixed target data: MSHT20 PDFs,'' Eur. Phys. J. C {\bf 81}, 341 (2021) [arXiv:2012.04684].

\end{thebibliography}
\end{document}